\documentclass[twoside]{article}
\usepackage{fleqn,espcrc2}

\newcommand{\agt}{\raise.3ex\hbox{$>$\kern-.75em\lower1ex\hbox{$\sim$}}~}
\newcommand{\alt}{\raise.3ex\hbox{$<$\kern-.75em\lower1ex\hbox{$\sim$}}~}

\newcommand{\AmS}{{\protect\the\textfont2
  A\kern-.1667em\lower.5ex\hbox{M}\kern-.125emS}}

\title{Diquark composites in the color superconducting phase of 
two flavor dense QCD}

\author{V.A. Miransky\address[BITP]{Bogolyubov 
Institute for Theoretical Physics, 252143, Kiev, 
Ukraine}\thanks{Present address: Department of Applied Mathematics,
University of Western Ontario, London, Ontario N6A 5B7, Canada.},
        I.A. Shovkovy\address[UC]{Physics Department, 
University of Cincinnati, Cincinnati, OH 45221-0011, 
USA}\address[UM]{School of Physics and Astronomy, 
University of Minnesota, Minneapolis, MN 55455, 
USA}\thanks{On leave of absence from Bogolyubov 
Institute for Theoretical Physics, 252143, Kiev, 
Ukraine.}, 
        L.C.R. Wijewardhana\addressmark[UC]}
      
\begin{document}

\begin{abstract}
We study the Bethe-Salpeter equations for spin zero diquark
composites in the color superconducting phase of $N_f=2$ cold dense
QCD. The explicit form of the spectrum of the diquarks, containing
an infinite tower of narrow (at high density) resonances, is
derived. It is argued that there are five pseudo-Nambu-Goldstone
bosons (pseudoscalars) that remain almost massless at large
chemical potential. These five pseudoscalars should play an important
role in the infrared dynamics of $N_f=2$  dense QCD.
\vspace{1pc}
\end{abstract}

\maketitle

In his studies Dmitrij Vasilievich Volkov was always led by the beauty
within the problems he considered. One of his passions was the theory
of spontaneous symmetry breaking, in particular, the dynamics of
Nambu-Goldstone (NG) particles (both bosons and fermions), to which
he contributed a great deal to our present understanding \cite{DVV1,DVV2}. 
Therefore we think it is most appropriate to report in this volume
a recent investigation of the dynamics of diquark composites (in
particular, diquark NG bosons) in cold dense QCD with two fermion flavors.

Only a few years ago, not much was known about the properties of
different phases in dense quark matter (see, however, 
Refs.~\cite{BarFra,Bail}). The situation drastically changed after 
the ground breaking estimates of the color superconducting order 
parameter were obtained in Refs.~\cite{W1,S1}. Within the framework 
of a phenomenological model, it was shown that the order parameter 
could be as large as 100 MeV. Afterwards, the same estimates  were
also obtained within the microscopic theory, quantum  chromodynamics
\cite{PR1,Son,us,SW2,PR2,H1,Brown1,us2}. The  further progress in the
field was mostly motivated by the  hope that the color
superconducting phase could be produced  either in heavy ion
experiment, or in the interior of neutron  (or rather quark) stars. 

Despite many advances \cite{CasGat,SonSt,Rho,HZB} in study of the
color superconducting phase of dense quark matter, the detailed 
spectrum of the diquark bound states (mesons) is still poorly
known.  In fact, most of the studies deal with the NG 
bosons of the three flavor QCD. At best, the indirect methods
of  Refs.~\cite{CasGat,SonSt,Rho,HZB} could probe the properties of
the  pseudo-NG bosons. It was argued in Ref.~\cite{us3}, however,
that, because of long-range interactions mediated by the gluons of
the magnetic type \cite{PR1,Son},  the presence of an infinite
tower of massive diquark states could  be the key signature of the
color superconducting phase of dense quark  matter. 

In this paper, we consider the problem of spin zero bound states in the
two flavor color superconductor using the Bethe-Salpeter (BS) equations.
We find that the spectrum contains five (nearly) massless states and an
infinite tower of massive singlets with respect to the unbroken
$SU(2)_{c}$ subgroup. Furthermore, in the hard dense loop improved ladder
approximation, the following mass formula is derived for the singlets:
\begin{equation}
M^{2}_{n} \simeq 4 |\Delta |^{2} 
\left(1-\frac{\alpha_{s}^{2}\kappa}{(2n+1)^{4}}\right),
~~n=1,2,\ldots,
\label{mass-singlet}
\end{equation}
where $\kappa$ is a constant of order 1 (we find that $\kappa\simeq
0.27$), $|\Delta |$ is the dynamical Majorana mass of quarks in the color
superconducting phase, and $\alpha_{s}$ is the value of the running
coupling constant related to the scale of the chemical potential
$\mu$.

At large chemical potential, we also notice an approximate
degeneracy between scalar and pseudoscalar channels. As a result 
of this parity doubling, the massive diquark states come in
pairs. In addition, there also exist five massless scalars and
five (nearly) massless pseudoscalars [a doublet, an antidoublet
and a singlet  under $SU(2)_{c}$]. While the scalars are removed
from the spectrum  of physical particles by the Higgs mechanism,
the pseudoscalars  remain in the spectrum, and they are the
relevant degrees of freedom  of the infrared dynamics. At 
high density, the massive and  (nearly) massless states are
narrow resonances.

In the case of two flavor dense QCD, the original gauge
symmetry  $SU(3)_{c}$  breaks down to the $SU(2)_{c}$ by Higgs
mechanism. The flavor $SU(2)_{L} \times SU(2)_{R}$ group remains
intact at the vacuum. The appropriate  order parameter is an
antitriplet in color and a singlet in flavor. Without loss of
generality, we assume that the order parameter points in the
third direction of the color space. In order to have a convenient
description of the bound states at the true vacuum, we introduce
the  following Majorana spinors,
\begin{eqnarray}
\Psi^{i}_{a} = \psi^{i}_{a}
+ \varepsilon_{3ab} \varepsilon^{ij} (\psi^{C})_{j}^{b} ,
&& a=1,2,  \label{Maj-psi}\\
\Phi^{i}_{a} = \phi^{i}_{a}
- \varepsilon_{3ab} \varepsilon^{ij} (\phi^{C})_{j}^{b} ,
&& a=1,2  ,\label{Maj-phi}
\end{eqnarray}
made of the Weyl spinors of the first two colors,
\begin{eqnarray}
\psi^{i}_{a}={\cal P}_{+} (\Psi_{D})^{i}_{a}, &\quad &
(\psi^{C})_{j}^{b}={\cal P}_{-} (\Psi_{D}^{C})_{j}^{b}, \\
\phi^{i}_{a}={\cal P}_{-} (\Psi_{D})^{i}_{a}, &\quad &
(\phi^{C})_{j}^{b}={\cal P}_{+} (\Psi_{D}^{C})_{j}^{b}.
\end{eqnarray}
Here $i,j=1,2$ are flavor indices,
${\cal P}_{\pm}=(1 \pm \gamma^5)/2$ are the left- and
right-handed projectors, $\Psi_{D}$ is the Dirac spinor, and 
$\Psi_{D}^{C}=C\bar{\Psi}^{T}_{D}$ is its charge conjugate. 
Regarding the quark of the third color, we use the Weyl spinors, 
$\psi^{i}$ and $\phi^{i}$, for left and right components, 
respectively (notice that the color index is omitted).

The BS wave functions of the bound diquark states 
in the channels of interest are given by
\begin{eqnarray}
(2\pi)^4 \hspace{-3.5mm}&&\hspace{-3.5mm}
\delta^4 (p_{+}-p_{-}-P)\mbox{\boldmath$\chi$}^{(\tilde{b})}_{a} (p,P) =
\nonumber \\ 
&& = \langle 0| T \Psi^{i}_{a}(p_{+}) \bar{\psi}_{i}(-p_{-})
|P; \tilde{b} \rangle_{L} , \label{def-chi} \\ 
(2\pi)^4 \hspace{-3.5mm}&&\hspace{-3.5mm}
\delta^4 (p_{+}-p_{-}-P)\mbox{\boldmath$\lambda$}_{(\tilde{a})}^{~b} 
(p,P)=
\nonumber \\
&& = \langle 0| T \psi^{i}(p_{+}) \bar{\Psi}_{i}^{b}(-p_{-})
|P; \tilde{a} \rangle_{L},  \label{def-lambda}\\
(2\pi)^4 \hspace{-3.5mm}&&\hspace{-3.5mm} 
\delta^4 (p_{+}-p_{-}-P)\mbox{\boldmath$\eta$} (p,P) = \nonumber \\
&& = \langle 0| T \Psi^{i}_{a}(p_{+})
\bar{\Psi}_{i}^{a}(-p_{-}) |P \rangle_{L},  \label{def-rho}\\
(2\pi)^4 \hspace{-3.5mm}&&\hspace{-3.5mm} 
\delta^4 (p_{+}-p_{-}-P)\mbox{\boldmath$\sigma$} (p,P) = \nonumber \\
&& = \langle 0| T \psi^{i}(p_{+})
\bar{\psi}_{i}(-p_{-}) |P \rangle_{L}, \label{def-sigma}
\end{eqnarray}
where $p=(p_{+}+p_{-})/2$ and the quantities on the right hand
side of these equations
are defined as the Fourier transforms of the corresponding
BS wave functions in the coordinate space.
There are also the BS wave functions constructed out of the right handed
fields $\Phi^{i}_{a}$ and $\phi^{i}$. One might notice that there
is another diquark channel, a triplet under $SU(2)_{c}$, that we
do not consider here. The reason is that the repulsion dominates
in such a channel, and  no bound states are expected [the triplet
comes from the  $SU(3)_{c}$ sextet].  

In order to derive the BS equations, we use the method developed
in Ref.~\cite{FGMS} for the case of zero chemical potential. To
this end, we need to know the quark propagators and the
quark-gluon interactions.

By introducing the multicomponent spinor that combines the Majorana 
spinors of the first two colors and the Weyl spinors of the third 
color, $\left( \Psi^{j}_{b} , \psi^{j} , \psi^{C}_{j} \right)^{T}$,
we find that the inverse propagator takes the following block-diagonal 
form:
\begin{equation}
G^{-1}_{p} = \mbox{diag} \left( 
S^{-1}_{p}\delta_{a}^{~b}\delta^{i}_{~j}, ~
s^{-1}_{p} \delta^{i}_{~j}, ~
\bar{s}^{-1}_{p} \delta_{i}^{~j} 
\right) , \label{propagator}
\end{equation}
where, upon neglecting the wave functions renormalization
of quarks \cite{Son,us,SW2,PR2,H1,Brown1,us2},
\begin{eqnarray}
 S^{-1}_{p}  \hspace{-3mm}&=&\hspace{-3mm}
-i\left( \not{\! p} +\mu \gamma^0 \gamma^5 + \Delta_{p} {\cal P}_{-}
+ \tilde{\Delta}_{p} {\cal P}_{+} \right) ,  \label{S_L} \\
 s^{-1}_{p}   \hspace{-3mm}&=&\hspace{-3mm} 
-i \left( \not{\! p} + \mu \gamma^0  \right) 
{\cal P}_{+},   \label{s_L} \\
 \bar{s}^{-1}_{p}  \hspace{-3mm}&=&\hspace{-3mm}
-i \left( \not{\! p} - \mu \gamma^0 \right) 
{\cal P}_{-}.  \label{s^C_L} 
\end{eqnarray}
Here the notation, $\Delta_{p} = \Delta^{+}_{p}
\Lambda^{+}_{p}  + \Delta^{-}_{p}\Lambda^{-}_{p}$,
$\tilde{\Delta}_{p}=\gamma^0 \Delta^{\dagger}_{p}\gamma^0$,
and $\Lambda^{\pm}_{p}=(1\pm \vec{\alpha}\cdot\vec{p}/|p|)/2$
are the same as in Ref.~\cite{us}. 

The bare vertex, $\gamma^{A\mu}$, is also a $3\times 3$ matrix,
\begin{equation}
\gamma^{A\mu} = \gamma^{\mu}
\left(\begin{array}{ccc}
\bar{\gamma}^{A}_{11} \delta^{i}_{~j}& 
\bar{\gamma}^{A}_{12} \delta^{i}_{~j}& 
\bar{\gamma}^{A}_{13} \varepsilon^{ij}\\
\bar{\gamma}^{A}_{21} \delta^{i}_{~j}& 
\bar{\gamma}^{A}_{22} \delta^{i}_{~j}& 
\bar{\gamma}^{A}_{23} \varepsilon^{ij}\\
\bar{\gamma}^{A}_{31} \varepsilon_{ij}& 
\bar{\gamma}^{A}_{32} \varepsilon_{ij}& 
\bar{\gamma}^{A}_{33} \delta_{i}^{~j}
\end{array}\right),
\end{equation}
with 
\begin{eqnarray}
\left(\bar{\gamma}^{A}_{11}\right)_{a}^{~b} &=& 
T^{Ab}_{a}-2\delta^{A}_{8} T^{8b}_{a} {\cal P}_{-},\\
\bar{\gamma}^{A}_{22} &=& T^{A3}_{3} {\cal P}_{+},\\
\bar{\gamma}^{A}_{33} &=& -T^{A3}_{3} {\cal P}_{-},\\
\left(\bar{\gamma}^{A}_{12} \right)_{a} &=&
T^{A3}_{a} {\cal P}_{+} ,\\
\left(\bar{\gamma}^{A}_{13}\right)_{a} &=&   
-\varepsilon_{3ac} T^{Ac}_{3} {\cal P}_{-},\\
\left(\bar{\gamma}^{A}_{21} \right)^{b} &=&
T^{Ab}_{3} {\cal P}_{+},\\
\left(\bar{\gamma}^{A}_{31} \right)^{b} &=&
-T^{A3}_{c} \varepsilon^{3cb} {\cal P}_{-},\\
\left(\bar{\gamma}^{A}_{23} \right)_{a} &=& 0,\\
\left(\bar{\gamma}^{A}_{32}\right)^{b} &=& 0,
\end{eqnarray}
where $T^{A}$ are the $SU(3)_{c}$ generators in the fundamental
representation. By making use of this vertex and the propagator in
Eq.~(\ref{propagator}), it is straightforward to derive the BS equations
in the (hard dense loop improved) ladder approximation. The details of the
derivation, as well as the explicit form of equations are given elsewhere
\cite{us-long}. Here we just note that the most transparent form of the
equations appears for the amputated BS wave functions, defined by
\begin{eqnarray}
\chi (p,P) = S^{-1}(p+\frac{P}{2}) \mbox{\boldmath$\chi$} 
(p,P) s^{-1}(p-\frac{P}{2}) , \label{chi-a} \\
\lambda (p,P) = s^{-1}(p+\frac{P}{2}) \mbox{\boldmath$\lambda$} 
(p,P) S^{-1}(p-\frac{P}{2}) , \label{lambda-a}\\
\eta (p,P) = S^{-1}(p+\frac{P}{2}) \mbox{\boldmath$\eta$} (p,P) 
S^{-1}(p-\frac{P}{2}) , \label{rho-a} \\
\sigma (p,P) = s^{-1}(p+\frac{P}{2}) \mbox{\boldmath$\sigma$} (p,P) 
s^{-1}(p-\frac{P}{2}) . \label{sigma-a}
\end{eqnarray}
In order to get a feeling of the problem at hand, let us 
briefly discuss the analysis of the BS equation for the 
$\chi$-doublet. In general, the BS wave function contains 
eight different Dirac structures \cite{footnote}. It is of 
great advantage to notice that only four of them survive in 
the center of mass frame, $P=(M_{b},\vec{0})$,
\begin{equation}
\chi_{a}^{(\tilde{b})} (p,0) = \delta_{a}^{~\tilde{b}} \hat{\chi}(p), 
\end{equation}
where 
\begin{eqnarray}
\hat{\chi}(p) = \left[\chi_{1}^{-} \Lambda^{+}_{p}
+ (p_{0}-\epsilon^{-}_{p}+\frac{M_{b}}{2}) 
\chi_{2}^{-} \gamma^0 \Lambda^{+}_{p} \right. \nonumber \\
\left. + \chi_{1}^{+} \Lambda^{-}_{p}
+ (p_{0}+\epsilon^{+}_{p}+\frac{M_{b}}{2}) 
\chi_{2}^{+} \gamma^0 \Lambda^{-}_{p} \right] {\cal P}_{+},
\label{ansatz} 
\end{eqnarray}
with $\epsilon^{\pm}_{p}=|\vec{p}|\pm \mu$ [the factors $(p_{0}\pm
\epsilon^{\pm}_{p}+M_{b}/2)$ are introduced here for convenience].
This is the most general structure that is allowed by the space-time
symmetries of the model. 

Now, in the particular case of the NG bosons, $M_{b}=0$, we will
show that the BS wave function is fixed by the Ward identities. 
Indeed, let us consider the following non-amputated vertex:
\begin{equation}
{\bf \Gamma}^{A,i}_{aj,\mu}(x,y) = \langle 0 | T 
j^{A}_{\mu}(0) \Psi^{i}_{a}(x) \bar{\psi}_{j}(y) 
| 0 \rangle , \label{cur-}
\end{equation}
where, for our purposes, it is sufficient to consider $A=4,\dots,8$
(that correspond to the five broken generators). In the (hard dense
loop improved) ladder approximation, the vertex satisfies the
following Ward identity \cite{us-long}:
\begin{equation}
P^{\mu} {\bf \Gamma}^{A,i}_{aj,\mu}(k+P,k) = i T^{A3}_{a}  
\delta^{i}_{j} \left[ s_{k} -S_{k+P} \right] {\cal P}_{-}.\label{27b} 
\end{equation}
As in the case of the BS wave functions, it is more convenient
to deal with the corresponding amputated quantity,
\begin{equation}
\Gamma^{A,i}_{aj,\mu}(k+P,k) = S^{-1}_{k+P}
{\bf \Gamma}^{A,i}_{aj,\mu}(k+P,k) s^{-1}_{k}.
\end{equation}
This latter satisfies the following identity:
\begin{equation}
P^{\mu} \Gamma^{A,i}_{aj,\mu}(k+P,k) = i T^{A3}_{a} \delta^{i}_{j} 
\left[ S^{-1}_{k+P} -s^{-1}_{k} \right] {\cal P}_{+}.
\label{ward} 
\end{equation}
By making use of the explicit form of the quark propagators in
Eqs.~(\ref{S_L}) and (\ref{s_L}), we could check that the right 
hand side of Eq.~(\ref{ward}) is non-zero in the limit $P\to 0$. 
This is possible only if the vertex on the left hand side 
develops a pole as $P\to 0$. After a simple calculation, we
obtain
\begin{equation}
\left. \Gamma^{A,i}_{aj,\mu}(k+P,k) \right|_{P\to 0} \simeq 
\frac{\tilde{P}^{\mu}}{P_{\nu}\tilde{P}^{\nu}} T^{A3}_{a} 
\delta^{i}_{j} \tilde{\Delta}_{k} {\cal P}_{+},
\label{pole} 
\end{equation}
where, we introduced $\tilde{P}^{\mu}=(P_{0},c_{\chi}^{2}\vec{P})$
with $c_{\chi}$ being the velocity of the NG boson in the 
$\chi$-doublet channel. 

By making use of the definition in Eqs.~(\ref{def-chi}) and
(\ref{chi-a}),  it is also not difficult to show that the pole
contribution to the  vertex function (\ref{pole}) is directly related
to the BS wave function. By omitting the details,
\begin{equation} 
\chi_{a}^{(\tilde{a})}(p,0) \equiv \delta_{a}^{(\tilde{a})} \chi(p,0)
=  \delta_{a}^{(\tilde{a})} \frac{\tilde{\Delta}_{p}}{F^{(\chi)}} 
{\cal P}_{+}, \label{chi-ward}
\end{equation}
where $F^{(\chi)}$ is the decay constant of the corresponding doublet
whose formal definition is given by
\begin{equation}
\langle 0| \sum_{A=4}^{7} T_{a}^{A3} j^{A}_{\mu} (0) | P, \tilde{b}
\rangle_{L} = i \delta_{a}^{\tilde{b}} \tilde{P}_{\mu} F^{(\chi)}.
\label{decay}
\end{equation}

By comparing the Dirac structures in Eqs.~(\ref{ansatz}) and 
(\ref{chi-ward}), we see that no components of the $\chi^{\pm}_{2}$
type appear in Eq.~(\ref{chi-ward}) which follows from the Ward 
identities. It was rewarding to establish that, in this approximation, 
the structure of the BS wave  function required
by the Ward identity is indeed a solution to the BS equation
for the $\chi$-doublet. A similar situation takes place for 
the $\eta$-singlet \cite{Brown2}. 

Now let us discuss the fate of the massless states that we obtain.
Altogether, there are five scalars and five pseudoscalars (a
doublet, an antidoublet and a singlet). Because of the Higgs
mechanism, the  scalars are removed from the spectrum.
Nevertheless, these scalar  bound states exist in the theory as
``ghosts" \cite{JJCN}, and one  cannot get rid of them
completely, unless a unitary gauge is found.  In fact, these
ghosts play a very important role in getting rid  of unphysical
poles from the on-shell scattering amplitudes  \cite{JJCN}.

As for the pseudoscalars, they remain in the spectrum as pseudo-NG
bosons. In the (hard dense loop improved) ladder approximation, they
look like NG bosons because the left and right sectors of quarks
decouple. One could  think of this as an effective enlargement of the
original color  symmetry from $SU(3)_{c}$ to  an approximate
$SU(3)_{c,L}\times SU(3)_{c,R}$. Then, since the approximate symmetry
of the ground state is $SU(2)_{c,L}\times SU(2)_{c,R}$, five scalar
NG bosons (which are removed by the Higgs mechanism) and five
pseudoscalar NG bosons (which remain in the spectrum) should appear.
Of course, in the full theory,  the pseudoscalars are only pseudo-NG
bosons. Indeed, they should get  non-zero masses due to higher orders
corrections that are beyond the improved ladder approximation
\cite{Wienberg}. Since the theory is weakly coupled at large chemical
potential, it is natural to expect that the masses of the pseudo-NG
bosons are small compared to the value  of the dynamical quark mass.

We conclude our discussion of the massless diquarks by emphasizing that
the low-energy dynamics of the two flavor QCD is dominated by massless
quarks of the third color (which might eventually get a small mass too if
another (non-scalar) condensate is generated \cite{W1,3-color}) and by the
five pseudoscalars that remain almost massless in the dense quark matter.
Of course, the gluons (glueballs) of the unbroken $SU(2)_{c}$ may also be
of some relevance but we do not study this question here.

Now, let us consider massive diquarks. The structure of the BS equations
becomes even more complicated in this case. In addition, one does not have
a rigorous argument to neglect the component functions like
$\chi^{\pm}_{2}$ in Eq.~(\ref{ansatz}). In spite of this, we argue that
all the approximations made before might still be reliable. Indeed, from
the experience of solving the gap equation (which coincides with the BS
equation for the massless states), we know that the most important region
of momenta in the integral equation is $|\Delta| \ll p \ll \mu$. In this
region, the kernel of the BS equations for massive states, $M_{b} \alt
|\Delta|$, is almost the same. The deviations appear only in the infrared
region where $p \alt |\Delta|$.

Therefore, in our analysis of the BS equations for massive states, we
closely follow the approximation used for the massless diquarks. By
assuming that the component functions depend only on the time component of
the momentum (compare with the analysis of the gap equation in
Refs.~\cite{Son,us,SW2,PR2,H1,Brown1,us2}), we arrive at the following
equation for the BS wave function of the massive singlets:
\begin{equation}
\eta^{-}_{1}(p) = \frac{\alpha_{s}}{4\pi} \int_{0}^{\Lambda} 
d q K^{(\eta)}(q) \eta^{-}_{1} (q) 
\ln \frac{\Lambda}{|q-p|}, \label{BS-mass}
\end{equation}
where $\Lambda=(4\pi)^{3/2}\mu/\alpha_{s}^{5/2}$, and the kernel
reads 
\begin{equation}
K^{(\eta)}(q) = \frac{\sqrt{q^{2}+|\Delta|^{2}}}
{ q^{2}+|\Delta|^{2} -\left(M_{\eta}/2\right)^{2} },
\label{K-eta}
\end{equation}
[$\eta^{-}_1(p)$ is a scalar function that appears in the decomposition of
the BS wave function $\eta$ over the Dirac matrices (compare with
Eq.~(\ref{ansatz})]. At this point it is appropriate to emphasize that the
Meissner effect plays an important role in the analysis of the massive
bound states. Indeed, our analysis shows that these massive states are
quasiclassical in nature, i.e., their binding energy is small compared to
the value of the gap [see Eq.~(\ref{mass-singlet})]. As a result, only the
long range interaction mediated by the unscreened gluons of the unbroken
$SU(2)_{c}$ is strong enough to produce these diquark states. We took this
into account in Eq.~(\ref{BS-mass}). For completeness, we mention that the
(nearly) massless (pseudo-) NG bosons are tightly bound, and the Meissner
effect is not so important for their binding dynamics.

By approximating the kernel (\ref{K-eta}) in each of the
following three regions: $0< q < \sqrt{|\Delta|^{2}
-\left(M_{\eta}/2\right)^{2} }$, $\sqrt{|\Delta|^{2}
-\left(M_{\eta}/2\right)^{2} } < q < |\Delta|$ and $|\Delta|< q
<\Lambda$, we could solve the BS equation (\ref{BS-mass})
analytically. Then, by matching the logarithmic derivatives of
the separate solutions, we obtain the spectrum of the massive
diquarks. By omitting the details, it is presented in
Eq.~(\ref{mass-singlet}).

We would like to emphasize that for large $\mu$ the hard dense
loop improved ladder approximation is reliable for the
description of those bound states. The point is that
a) the region of momenta primarly responsible for the formation
of these composites is $E_{bind} \alt q \ll \mu$, where the binding
energy $E_{bind}\sim \alpha_{s}^{2}\Delta$, and b) $E_{bind}\to\infty$  
as $\mu\to\infty$ \cite{Son,us,SW2,PR2}.
Therefore the vacuum effects 
are higher order ones in $\alpha_{s}$ in that region. Because of
that,
the hard dense loop improved ladder approximation, in which
the contribution of the vacuum effects to the running of the
coupling constant is neglected and only the running due to the 
polarization effects provided by the quark matter (non-zero $\mu$)
is taken into account, is justifiable for large $\mu$.

Now, let us consider the case of massive diquarks in the doublet
channel. As is easy to check, the binding interaction in this
channel is exclusively due to the five gluons affected by the
Meissner effect. The approximate BS equation looks similar to
Eq.~(\ref{BS-mass}), but with a different kernel and $|q-p|$ in
the logarithm replaced by $|q-p|+c\Delta$ where $c=O(1)$  is a
constant \cite{us}. At high density when the coupling constant
is weak, this equation does not allow a non-trivial solution for
$M\neq 0$. From the physical viewpoint, this indicates that the
heavy gluons, with $M_{gl}\sim (\alpha_{s} \mu^2 \Delta)^{1/3}\gg
\Delta$, cannot provide a sufficiently strong attraction to form
massive radial excitations of the NG and pseudo-NG bosons. 

At the end, let us note that the massive diquark states may truly be just
resonances in the full theory, since they could decay into the pseudo-NG
bosons and/or gluons (glueballs) of the unbroken $SU(2)_{c}$. At high
density, however, both the running coupling $\alpha_{s}(\mu)$ and the
effective Yukawa coupling $g_{Y} = |\Delta|/F \sim |\Delta|/\mu$
\cite{SonSt,Rho,HZB,Rischke}) are small, and, therefore, these
massive resonances are narrow.

In conclusion, in this paper we studied the problem of diquark
bound states in the color superconducting phase  of $N_f=2$
dense QCD. While the scalar NG bosons  are ghosts in the theory,
the pseudoscalar pseudo-NG  bosons are physical particles that
should play an important role in the infrared. We also obtained
the spectrum of the massive narrow diquark resonances, whose 
existence would be a clear signature of the unscreened long
range forces in dense QCD.

{\bf Acknowledgments}.
V.A.M. thanks V.P. Gusynin for useful discussions. The work of V.A.M.  
and I.A.S. was partly supported by the Grant-in-Aid of Japan Society for
the Promotion of Science No. 11695030. The work of I.A.S. was supported by
the U.S. Department of Energy Grants No.~DE-FG02-84ER40153 and
No.~DE-FG02-87ER40328. The work of L.C.R.W. was supported by the U.S.  
Department of Energy Grant No.~DE-FG02-84ER40153.


\begin{thebibliography}{99}
\bibitem{footnote} Note that in the case of zero chemical
potential there are only four different Dirac structures,
see Ref.~\cite{FGMS}.
\end{thebibliography}
\end{document}